\documentstyle{aipproc}

\begin{document}

\title{Four Testable Predictions \\ of Instanton Cosmology}

\author{R.E. Allen}

\address{Center for Theoretical Physics, Texas A\&M University\\
College Station, Texas 77843, USA}

\maketitle
\begin{abstract}
A new cosmological model makes the following predictions: (1) The
deceleration parameter $q_{0}$ is approximately zero. (2) The mass
density parameter $\Omega_{m}$ is less than 1.
(3) The universe is spatially closed, but is
asymptotically flat as $t\rightarrow \infty $, regardless of its
matter content. (4) The age of the universe is approximately 15 Gyr
if the Hubble parameter $h$ is approximately 0.65.
\end{abstract}

\address{Center for Theoretical Physics, Texas A\&M University\\
College Station, Texas 77843}

\address{Center for Theoretical Physics, Texas A\&M University\\
College Station, Texas 77843}

In a new theory~\cite{a}, the evolution of the cosmic scale factor
$R\left( t\right) $
is determined by both the Einstein field equations and an SU(2) cosmological
instanton. The instanton dominates in the later universe, making
$R\left(t\right)$ approximately proportional to $t$:
\begin{equation}
dR/dt=K \quad ,\quad t \rightarrow \infty
\end{equation}
where $K$ is a constant.
On the other hand, Einstein gravity dominates in the early universe,
giving the usual Friedmann equation
\begin{equation}
\left( dR/dt\right) ^{2}+k=\left( 8\pi G/3\right) \rho R^{2}\quad ,\quad
t \rightarrow 0
\end{equation}
with $k$ positive (since the universe is spatially a 3-sphere).
Suppose that a crossover time $t_{c}$ is defined by
\begin{equation}
\left( 8\pi G/3\right) \rho R^{2}=K^{2}\quad ,\quad t=t_{c}.
\end{equation}
Since $\rho R^{2}$ decreases with $R$ and $t$, we then have
\begin{equation}
\rho <\left( \frac{3}{8\pi G}\right) \frac{K^{2}}{R^{2}}<\left( \frac{3}
{8\pi G}\right) \left( \frac{dR/dt}{R}\right) ^{2}\quad ,\quad t\gg t_{c}
\end{equation}
or
\begin{equation}
\Omega _{m}<1\quad ,\quad t\gg t_{c}
\end{equation}
where
\begin{equation}
\Omega _{m}\equiv \frac{\rho }{\rho _{c}}\quad ,\quad \rho _{c}\equiv \left(
\frac{3}{8\pi G}\right) H^{2}\quad ,\quad H\equiv \frac{dR/dt}{R}.
\end{equation}
According to (1), we also have
\begin{equation}
q\equiv -\frac{d^{2}R/dt^{2}}{H^{2}R}\approx 0\quad ,\quad t\gg t_{c}.
\end{equation}

With the reasonable assumption that the age of the universe is
substantially larger than $t_{c}$, the present theory thus makes
several testable predictions:

\bigskip
{\bf (1) The current deceleration parameter} ${\bf q_{0}}$ {\bf is
approximately zero.}

\bigskip
{\bf (2) The current mass density parameter} ${\bf \Omega }_{m}$
{\bf is less than 1.}

\bigskip
Two additional predictions are obtained below:

\bigskip
{\bf (3) The universe is spatially closed, but is asymptotically flat
as }${\bf t\rightarrow \infty ,}$ {\bf regardless of its matter content.}

\bigskip
{\bf (4) The age of the universe is approximately 15 Gyr if the Hubble
parameter }${\bf h}${\bf \ is approximately 0.65.}

\bigskip Let us consider the origin of these results in more detail,
beginning with the epoch $t\gg t_{c}$ where the cosmological instanton
dominates. In the theory of Ref. 1, this instanton is a topological defect
in a GUT Higgs field which condenses in the very early universe. Its order
parameter $\Psi _{s}$ has the symmetry group SU(2)$\times $U(1)$\times $
SO(10). The SO(10) gauge symmetry can be ignored in the present context,
however, so $\Psi _{s}$ may be regarded as simply an SU(2)$\times $U(1)
order parameter in four-dimensional spacetime. With $x^{0}$ chosen to be the
radial coordinate, there is a singularity at $x^{0}=0$ which is interpreted
as the big bang. This singularity is physically admissible because it is
perfectly analogous to the singularity at the center of a superfluid vortex.
One can in fact define a ``superfluid velocity''
\begin{equation}
v^{\mu }=-im^{-1}U^{-1}\partial ^{\mu }U \quad \mbox{where} \quad
\Psi _{s}=n_{s}^{1/2}U\eta _{0}
\end{equation}
and $\eta _{0}$ is a constant 2-component vector. The field
$e_{\alpha}^{\mu }$ defined by
\begin{equation}
e_{\alpha }^{\mu }=v_{\alpha }^{\mu } \quad , \quad
v^{\mu }=v_{\alpha }^{\mu }\sigma ^{\alpha }
\end{equation}
is interpreted as the vierbein which determines the geometry of
spacetime~\cite{a}.

\bigskip
If $a$ is the strength of the
cosmological instanton centered on $x^{0}=0,$ the constraint of fixed
SU(2) topological charge requires that~\cite{a,a1,ordinary}
\begin{equation}
e_{\alpha }^{\mu }=\delta _{\alpha }^{\mu }a/mx^{0} \quad , \quad
e_{\mu }^{\alpha }=\delta _{\mu }^{\alpha }mx^{0}/a
\end{equation}
In addition, larger values of $x^{0}$ correspond to larger
3-spheres~\cite{a}, so
$dx^{k}=x^{0}d\overline{x}^{k}$, where $d\overline{x}^{k}$ represents a
distance within the unit 3-sphere. There are thus two contributions to the
expansion of the universe in the present model, with both the 3-space metric
tensor $g_{kl}$ and the separation of comoving points increasing as
functions of $x^{0}$.

\bigskip The vierbein of (10) leads to a metric with the
Robertson-Walker form
\begin{eqnarray}
ds^{2} &=&\eta _{\alpha \beta }\,e_{\mu }^{\alpha }e_{\nu }^{\beta
}\,dx^{\mu }\,dx^{\nu } \\
&=&-\left( \frac{mx^{0}}{a}\right) ^{2}\left( dx^{0}\right) ^{2}+\left[
\frac{m\left( x^{0}\right) ^{2}}{a}\right] ^{2}d\overline{x}^{k}d\overline{x}
^{k} \\
&=&-dt^{2}+R\left( t\right) ^{2}\left( \frac{dr^{2}}{1-r^{2}}+r^{2}d\Omega
\right)
\end{eqnarray}
where $\eta _{\alpha \beta }=diag\left( -1,1,1,1\right) $ is the Minkowski
metric tensor and
\begin{equation}
R\left( t\right) =\left( m/a\right) \left( x^{0}\right) ^{2}\quad \quad
,\quad \quad dt=\left( m/2a\right) d\left( x^{0}\right) ^{2}.
\end{equation}
Then (1) follows immediately, with $K=2$.

\bigskip In the later universe, therefore, 
$R\left( t\right) $ is primarily determined
by the topology of the cosmological instanton. The mass density $\rho$ 
has some influence at any finite time $t$, however, and the {\it
local} metric tensor near a massive object is determined by
the Einstein field equations to an extremely good approximation.  As usual,
these equations follow from the variational principle $\delta S_{L}/\delta
g^{\mu \nu }=0$ if there are no constraints, where $S_{L}$ is the full
Lorentzian action~\cite{a}. In the present theory, there is a constraint
imposed
by topology on a cosmological scale, but this constraint is of negligible
importance on, e.g., a planetary scale. Near a dense concentration of
mass in our highly heterogeneous universe,
the local deformation of spacetime geometry is still accurately determined
by the usual equations of general relativity.

\bigskip Furthermore, the Einstein field equations are also a very
good approximation in the early universe, since the instanton contribution
of (1) to $\left( dR/dt\right) ^{2}$ is constant, whereas the usual
gravitational contribution of (2) grows as $R\rightarrow 0$, with $\rho
\propto $ $R^{-3}$ (in the matter-dominated era) or $R^{-4}$ (in the
radiation-dominated era).

\bigskip
The spatial curvature associated with (13) is proportional to
$R\left( t\right) ^{-2}$~\cite{misner}. The present model thus requires
a universe that is asymptotically flat,
because the topology of the cosmological instanton requires that $R\left(
t\right) \propto t$ for $t\rightarrow \infty $. This is true regardless of
the value of $\Omega _{m}$. Observations of the cosmic microwave background
and of large scale structure indicate that the universe is very nearly flat
but that $\Omega _{m}\approx 0.3$~\cite{webster}.

\bigskip For $t\gg t_{c}$, (1) implies that $H^{-1}\approx t$
rather than $\frac{3}{2}t$. The age of
the universe is then given by $t_{0}\approx H_{0}^{-1}$ rather than
$\frac{2}{3}
H_{0}^{-1}$. With the conventional definition $H_{0}=100h$ km s$^{-1}$Mpc$
^{-1}$, a value of $h=0.65$~\cite{branch,freedman}
implies that $t_{0}\approx 15$ Gyr rather than $10$
Gyr. This larger value is more consistent with recent estimates of $10-14$
Gyr for the ages of the oldest globular clusters~\cite{chaboyer}.

\bigskip  The present theory thus makes a number of testable
predictions. In particular, the prediction
$q_{0}\approx 0$~\cite{coasting} clearly demarcates the present theory
from standard cosmology with a flat spacetime,
in which $q_{0}=1/2$~\cite{peebles,kolb}, and from $\Lambda $CDM models,
in which $q_{0}\approx -1/2$~\cite{turner}.
Quantitative measurements of $q_{0}$ are now becoming
attainable~\cite{perlmutter,garnavich}, so it should be possible
to subject this prediction to convincing tests in the
near future.

\bigskip

\begin{center}
{\large \bf Acknowledgement}
\end{center}
This work was supported by the Robert A. Welch Foundation.

\end{document}